\documentstyle[eqsecnum,aps]{revtex}
\def\mbf#1{\setbox0=\hbox{$#1$}%
     \kern-.4pt\copy0\kern-\wd0
     \kern.2pt\copy0\kern-\wd0
     \kern.2pt\copy0\kern-\wd0
     \kern.2pt\copy0\kern-\wd0
     \kern.2pt\copy0
}
\def\mbff#1{\setbox0=\hbox{$#1$}%
     \kern-.6pt\copy0\kern-\wd0
     \kern.2pt\copy0\kern-\wd0
     \kern.2pt\copy0\kern-\wd0
     \kern.2pt\copy0\kern-\wd0
     \kern.2pt\copy0\kern-\wd0
     \kern.2pt\copy0\kern-\wd0
     \kern.2pt\copy0
}
\def\ie{\mbox{i.e.}}

\def\dA{{A'}}
\def\dB{{B'}}
\def\dC{{C'}}

\def\dm{{m'}}
\def\eps{\epsilon}

\def\sig{\sigma}

\def\half{{1 \over 2}}
\def\halfi{{i \over 2}}
\def\thirdh{{3 \over 2}}

\def\bo{{\bar o}}

\def\balpha{{\bar \alpha}}
\def\bbeta{{\bar \beta}}
\def\biota{{\bar \iota}}
\def\bzeta{{\bar \zeta}}

\def\bpsi{{\bar \psi}}

\def\Mp{{M_p}}
\def\hF{{\hat F}}
\def\bfD{\mbf{D}}
\def\bfF{\mbf{F}}
\def\bfR{\mbf{R}}

\def\bfbeta{\mbf{\beta}}
\def\bfgamma{\mbff{\gamma}}

\def\bfepsilon{\mbff{\epsilon}}

\def\bfphi{\mbf{\phi}}

\def\bfpsi{\mbf{\psi}}
\def\bfomega{\mbff{\omega}}
\mathchardef\Lag="724C
\def\deri{\partial}
\def\dzero{0'}
\def\done{1'}
\def\pinfin{{(+\infty,in)}}
\def\pinfout{{(+\infty,out)}}
\def\minfin{{(-\infty,in)}}
\def\minfout{{(-\infty,out)}}
\def\rslash{\partial\kern-0.026em\raise0.17ex\llap{/}%
\kern0.026em\relax}
\def\Dslash{D\kern-0.15em\raise0.17ex\llap{/}\kern0.15em\relax}

\begin{document}
\draft
\title{
	On scattering off the extreme Reissner-Nordstr\"om 
	black hole \\ in $N=2$ supergravity 
}
\author{
	Takashi Okamura 
	\thanks{email address: okamura@th.phys.titech.ac.jp}
}
\address{
	Department of Physics, Tokyo Institute of Technology, \\
	Oh-okayama, Meguro, Tokyo 152, Japan
}

\date{April 1, 1997}
\maketitle
\begin{abstract}
The scattering amplitudes for the perturbed fields of 
the $N=2$ supergravity 
about the extreme Reissner-Nordstr\"om black hole
is examined.
Owing to the fact that the extreme hole is a BPS state of the theory
and preserves an unbroken global supersymmetry($N=1$),
the scattering amplitudes of the component fields 
should be related to each other.
In this paper, we derive the formula of the transformation of 
the scattering amplitudes.
\end{abstract}
\section{Introduction}
Solitons that are non-perturbative configurations play 
an important role for studying non-perturbative aspects 
of quantum field theories.
A soliton is a classical solution which is stationary,
regular and classically and quantum mechanically stable configuration
with finite localized energy.
Solitons often have some conserved charges.
From the stability of the configurations 
in classical and quantum theory, we may think a soliton 
as the least energy state whose energy is given by the charges.
Further we may expect the inequality between the mass and 
charges of a soliton.
In fact, we have the inequality in supersymmetric theories and
call saturated states BPS states
\cite{WittenOlive}.
Although BPS states are massive and break the supersymmetry,
they still have some unbroken supersymmetry.

In the Einstein-Maxwell theory, 
extreme Reissner-Nordstr\"om solutions behave 
as gravitational solitons
\cite{Hajicek,Gibbons}.
The Einstein and the Einstein-Maxwell system can be 
embedded in supergravity theories.
In the asymptotically flat spacetime, we can obtain the global charges
that generate the rigid supersymmetry
\cite{algebra}.
Therefore we can follow the argument in the rigid supersymmetry formally.
For the $N=1$ supergravity the positivity of energy is 
suggested
\cite{algebra,positiveEsugra}.
Subsequently Witten established the positivity for 
the general relativity using the trick of \lq\lq Witten spinor"
\cite{positiveEeinstein}.
Further, using the \lq\lq Witten spinor" motivated 
by the transformation law of gravitini in the $N=2$ supergravity, 
Gibbons and Hull 
\cite{GibbonsHull}
established the inequality between the mass and 
the electromagnetic charges.
Further they showed that the saturated configurations are 
Majumdar-Papapetrou(MP) solutions,
which are assemblages of the extreme Reissner-Nordstr\"om holes,
and that the MP solutions have unbroken supersymmetries.
More generalizations of their results are available in Refs.
\cite{KalloshLindeETAL,GibbonsKastorETAL}.

In addition, the non-renormalization theorem of on-shell effective
action for the MP solutions was established
\cite{KalloshLindeETAL,Kallosh}.
Although supergravity has better ultraviolet behavior than 
the general relativity, it is known that 
supergravity is non-renormalizable at the perturvative level 
and is not regarded as the final theory.
However we may expect that the final theory should 
include supergravity and may think the non-renormalization 
for the MP solutions as a guiding principle to the final theory.

Thus, it is very significant to investigate extreme black holes 
in classical and semiclassical framework through the general relativity and 
supergravity.
To more understand extreme holes that are regarded as a kind of 
\lq\lq vacuum" state, we need to study the fluctuation(excitation) about
them.

Originally, the study of the perturbation about 
the extreme Reissner-Nordstr\"om hole was motivated by
the interest in the no hair conjecture in supergravity
\cite{AichelburgGuven}
and by the interpretation problem of the paradoxical thermal properties
of extreme-dilaton black holes
\cite{HolzheyWilczek}.

Recently in the study on the method of calculating 
quasinormal frequencies of the extreme Reissner-Nordst\"om hole,
Onozawa, et.al. numerically
\cite{oursI}
found that the quasinormal frequencies of 
gravitational waves and electromagnetic waves about it
coincide by a suitable shift of the angular momentum indices.
Due to the fact that quasinormal frequency is resonance pole
of scattering wave, we may expect that gravitational and electromagnetic
wave have the same reflection and transmition amplitude.
Subsequently, they established coincidence between S-matrices of
perturbations of gravitational, electromagnetic
and spin-$3/2$ fields(gravitini) about 
the extreme Reissner-Nordstr\"om background in the $N=2$ supergravity
by finding relation between the Regge-Wheeler potentials of perturbations
\cite{oursII}.
From the fact that the extreme Reissner-Nordst\"om hole 
is a BPS state in extended supergravity,
we expect that the coincidence is related to the fact that 
BPS states preserve unbroken supersymmetries.
The purpose of this paper is to derive the relation between
scattering matrices of graviton, gravitini and photon 
using supersymmetric transformation.

The paper is organized as follows:
In Sec. II we briefly review the perturbation equations through
the Newman-Penrose formalism and the scattering problem.
In Sec. III we give the supersymmetric transformation law 
between the curvatures of perturbed fields.
In Sec. IV we seek the correspondence between 
the radial parts of the perturbations with the suitable total angular
momentum and the relations of the reflection and transmition
coefficients for them.
Sec. V is devoted to summary.
\section{The perturbation equations}
By linearizing the $N=2$ supergravity 
\cite{sugra}
about a purely bosonic background, 
we have the perturbation equations
for linearized Einstein-Maxwell system and for 
linearized O(2) doublet of spin-$3/2$ fields, which they are decoupled.
Here we follow Chandrasekhar 
\cite{Chandra}
and Torres del Castillo \& Silva-Ortigoza 
\cite{TorresETAL}
for bosonic perturbations and fermionic ones respectively.
Hereafter we adopt unit $M=1$, where $M$ is the mass of 
background black hole. 

On the background of the Reissner-Nordstr\"om solution,
the bosonic perturbations are described by 
the Regge-Wheeler equation and the fermionic ones
by the similar equation thorough the Newmann-Penrose formalism.

For the perturbations with the helicity-($+1, +\thirdh, +2$), 
$Y_{+s}$($s=1,\thirdh,2$), in phantom gauge
\cite{phantom},
their equations of radial parts
are
\begin{eqnarray}
	& &\Lambda^2 Y_{+s}+P_s \Lambda_- Y_{+s} - Q_s Y_{+s}=0~,
\label{eq:eqm} \\
	& &{d \over dr_*}={\Delta \over r^2}{d \over dr}~,
\qquad\qquad \Delta=r^2-2r+Q^2~,
\label{eq:defDelta} \\
	& &\Lambda_{\pm} \equiv {d \over dr_*} \pm i \Omega~,
\qquad\qquad \Lambda^2 \equiv \Lambda_+ \Lambda_-~,
\label{eq:defLambda}
\end{eqnarray}
where $r_*$ is the tortoise coordinate and we omit the index of 
distinguishing two gravitini because they follows the same equation
as expected from O(2) symmetry between them.

The each $P_s$ and $Q_s$ are given by, for $s=1,2$,
\begin{mathletters}
\begin{eqnarray}
	& &P_s \equiv {d \over dr_*}\ln \Bigl({r^8 \over D_s} \Bigr)~,
\qquad
	D_s \equiv \Delta^2 \Bigl(1+{2 q_s \over \mu^2 r} \Bigr)~,
\label{eq:defbosonPsDs} \\
	& &Q_s \equiv \mu^2{\Delta \over r^4}
	\Bigl(1+{2 q_s \over \mu^2 r} \Bigr)
	\Bigl(1+{ q_{s'} \over \mu^2 r} \Bigr) 
	\qquad (s,s'=1,2~;~s \ne s')~,
\label{eq:defbosonQs}
\end{eqnarray}
\end{mathletters}
where $q_{1,2}$ are defined by
\begin{equation}
	q_1=3+\sqrt{9+4 Q^2 \mu^2}~, \qquad\qquad\quad
	 q_2=3-\sqrt{9+4 Q^2 \mu^2}~,
\label{eq:defqI}
\end{equation}
and for $s=\thirdh$,
\begin{mathletters}
\begin{eqnarray}
	P_{\thirdh} &\equiv &{3 \over r^3}(r^2-3r+2 Q^2)~,
\label{eq:deffermiPs} \\
	Q_{\thirdh} &\equiv & {\Delta \over r^6}
	(\lambda r^2+2r-2Q^2)~.
\label{eq:deffermiQs}
\end{eqnarray}
\end{mathletters}

The radial perturbations $Y_{+s}$ are constructed in two ways.
One is by the perturbed Weyl scalar $\Psi_0$ and the perturbed 
spin connection $\kappa$ as
\begin{eqnarray}
	& &Y_{+s}(r)={\Delta^2 \over r^3}F_{+s}(r)~,
\qquad\qquad\qquad
	F_{+s}=R_{+2}(r)+{q_s k(r) \over \mu}~,
\label{eq:defYF} \\
        & &\Psi_0=R_{+2}(r)S_{+2}(\theta) e^{i(\Omega t+m \phi)}~,
\quad 
	\kappa=\sqrt{2} r^2 k(r)S_{+1}(\theta) e^{i(\Omega t+m \phi)}~,
\label{eq:defRk}
\end{eqnarray}
where the constant $\mu$ is a eigenvalue of 
the spin-weighted spherical harmonics,
\begin{eqnarray}
        & &\Lag^{\dagger}_{-1}\Lag_2  S_{+2}=-\mu^2 S_{+2}~,
        \qquad\qquad
        \Lag_2\Lag^{\dagger}_{-1} S_{+1}=-\mu^2 S_{+1}~,
\\
	& &\mu=\sqrt{(J-1)(J+2)}~.
\label{eq:spherical}
\end{eqnarray}
The operators $\Lag_n$ and $\Lag^{\dagger}_n$ are defined by
\begin{mathletters}
\label{mlet:defLag}
\begin{eqnarray}
	\Lag_n &\equiv & \partial_\theta+{m \over \sin\theta}
	+n \cot\theta~,
\label{eq:defLagop} \\
	\Lag^{\dagger}_n &\equiv & \partial_\theta-{m \over \sin\theta}
	+n \cot\theta~.
\label{eq:defLagopdag}
\end{eqnarray} 
\end{mathletters}
Besides, the functions $S_{+1}$ and $S_{+2}$ are related in the manner
\begin{equation}
	\Lag_2 S_{+2}=\mu S_{+1}~,
        \qquad\qquad\qquad\qquad
        \Lag^{\dagger}_{-1} S_{+1}=-\mu S_{+2}~.
\label{eq:sphrelation}
\end{equation}

Another is by the Weyl scalar $\Psi_1$ and 
the spin connection $\sig$ as
\begin{eqnarray}
	& &G_{+s}(r)=R_{+1}(r)+{q_s \over \mu}~s(r) ~,
\label{eq:defG} \\
        & &\Psi_1={1 \over r \sqrt{2}}
        R_{+1}(r)S_{+1}(\theta) e^{i(\Omega t+m \phi)}~,
\, 
        \sigma= r s(r)S_{+2}(\theta) e^{i(\Omega t+m \phi)}~,
\label{eq:defRs}
\end{eqnarray}
and $G_{+s}$ are related to $Y_{+s}$ through the relations,
\begin{equation}
	\Delta \Bigl(D^{\dagger}_2-{3 \over r} \Bigr)F_{+s}
	=\mu \Bigl(1+{2 q_s \over \mu^2 r} \Bigr)G_{+s'}
	\qquad (s,s'=1,2;~s \ne s')~,
\label{eq:relGF}
\end{equation}
where the operator $D_n$ is
\begin{equation}
	D_n \equiv \partial_r+{i r^2 \Omega \over \Delta}
	+2n{r-1 \over \Delta}~, \qquad D^{\dagger}_n=(D_n)^*~.
\label{defDop}
\end{equation}

For the helicity-($+3/2$) perturbations,
the supersymmetric gauge invariant quantities are constructed by
the supercovariant curvature of the spin-$3/2$ fields $\psi^i_\mu$ as
\begin{eqnarray}
	& &H^i_0 =\Psi^i_{(ABC)}o^A o^B o^C~,
\label{eq:defHoI} \\
	& &\Psi^{i~A}{}_{BC}=\half \Bigl[~
	\bfD_{(B|\dA|}\psi^{iA}{}_{C)}{}^{\dA}
        +{i \over \Mp}\eps^{ij}~\bfF^A {}_{(B} 
	\bpsi^{j\dA}{}_{|\dA| C)} \Bigr]~,
\label{eq:defPsiABC}
\end{eqnarray}
where $\bfD_\mu$ is covariant derivative with respect to 
the spin connection 
$\bfomega_{\mu ab}=\bfomega_{\mu AB}\eps_{\dA\dB}
+\bar{\bfomega}_{\mu \dA\dB}\eps_{AB}$, 
for example,
\begin{equation}
	\bfD_\mu \eta_A = \partial_\mu \eta_A 
	+ \bfomega_{\mu A}{}^B \eta_B~,
\label{eq:defCDI}
\end{equation}
and the bold-face letters indicate the background quantities, 
$o^A$ is a principal spinor of $\bfF_{AB}$, and $\bfF_{AB}$
is a 2-spinor representation of the self-dual part of 
electromagnetic field strength $\bfF_{\mu\nu}$.
Hence we obtain the modes of the helicity-($+3/2$) perturbations,
\begin{eqnarray}
	& &Y^j_{+\thirdh}={\Delta^{\thirdh} \over r^2}R^j_{+\thirdh}~,
\\
	& &H^j_0=R^j_{+\thirdh}(r)S_{+\thirdh}(\theta)
	~e^{i(\Omega t+m \phi)}~,
\label{eq:defYthirdh}
\end{eqnarray}
where $m$ is $+1/2$ or $-1/2$ and 
the spin-weight $+3/2$ spherical harmonics $S_{+\thirdh}$ 
satisfies
\begin{eqnarray}
	& &\Lag^{\dagger}_{-\half}\Lag_{\thirdh}S_{+\thirdh}
	=-\lambda S_{+\thirdh}~,
\\
	& &\lambda=(J_s-\half)(J_s+\thirdh)
	 \qquad\qquad (J_s=\thirdh, {5 \over 2},...)~.
\label{eq:sphthirdh}
\end{eqnarray}

To set the scattering problem, we need the normalized in(out)-going
wave forms for $Y_{+s}$ 
at the asymptotic regions($r_* \rightarrow \pm\infty$).
At $r_* \rightarrow \infty$, its asymptotic form of each
normalized perturbation $Y_{+s}$($s=1,\thirdh,2$) become
\begin{equation}
	Y^{\pinfin}_{+s} \sim -4 \Omega^2 e^{+i \Omega r_*} 
\qquad \hbox{and} \qquad 
	Y^{\pinfout}_{+s} \sim -{K_s \over 4 \Omega^2 r^{2 \Vert s \Vert} }
	e^{-i \Omega r_*}~,
\label{eq:asymYsp}
\end{equation}
where $\Vert s \Vert=2$ for $s=1,2$ and $\Vert s \Vert=\thirdh$ 
for $s=\thirdh$.
$K_s$ are defined by
\begin{eqnarray}
	& &K_s \equiv \mu^2 (\mu^2 +2)+2 i \Omega \beta_s~,
\qquad \beta_s^2 \equiv q^2_{s'} \qquad (s,s'=1,2; s \ne s')~,
\label{eq:defKs} \\
	& &K_{\thirdh} \equiv 2 i \Omega (\kappa_{\thirdh}
	+2 i \Omega \beta_{\thirdh})~,
\label{eq:defKthirdh} \\
	& &\kappa^2_{\thirdh} \equiv \Bigl[ \Bigl(J_s-\half \Bigr)
	\Bigl(J_s+\half \Bigr) \Bigl(J_s+\thirdh \Bigr) \Bigr]^2~,
\qquad	\beta^2_{\thirdh} \equiv 4~.
\label{eq:defkappabetathirdh}
\end{eqnarray}
Similarly, at $r_* \rightarrow -\infty$ for $s=1,2$
\begin{eqnarray}
	& &Y^{\minfout}_{+s} \sim 4 i \Omega  \Bigl(
	i \Omega -{r_+ -Q^2 \over r_+^3} \Bigr)
	\exp \left(+i \Omega r_* \right) 
\nonumber \\
	& &Y^{\minfin}_{+s} \sim 
	{\displaystyle{K_s (1+{2 q_s \over \mu^2 r_+})
	\Delta^{\Vert s \Vert}} \over 
	\displaystyle{ 4 r_+^{4 \Vert s \Vert} 
	\Bigl(i \Omega - 2{r_+ -1 \over r_+^4} \Bigr)
	\Bigl(i \Omega - {r_+ -Q^2 \over r_+^3} \Bigr)} }
	 \exp \left(-i \Omega r_* \right)~,
\label{eq:asymYsm}
\end{eqnarray}
and for $s=\thirdh$
\begin{eqnarray}
	& &Y^{\minfout}_{+\thirdh} \sim 4 i \Omega  \Bigl(
	i \Omega -{r_+ -1 \over 2 r_+^2} \Bigr)
	\exp \left(+i \Omega r_* \right) 
\nonumber \\
	& &Y^{\minfin}_{+\thirdh} \sim 
	{K_s  \Delta^{\Vert s \Vert} \over 
	\displaystyle{ 
	4 r_+^{4 \Vert s \Vert} \Bigl(i \Omega - {r_+ -1 \over 2 r_+^2} \Bigr)
	\Bigl(i \Omega - {3 \over 4}{r_+ -1 \over r_+^2} \Bigr)} } 
	\exp \left(-i \Omega r_* \right)~.
\label{eq:asymthirdh}
\end{eqnarray}

Using the above basis, the scattering problems of the perturbations 
are set as
\begin{eqnarray}
	Y_{+s} &\sim & Y^{\pinfin}_{+s} 
	+ R_s (\Omega) Y^{\pinfout}_{+s} 
		\qquad (r_* \rightarrow \infty)~,
\nonumber \\
	&\sim & \qquad\qquad\qquad T_s (\Omega) Y^{\minfout}_{+s} 
		\qquad (r_* \rightarrow -\infty)~,
\label{eq:asymYs}
\end{eqnarray}
where $R_s$ and $T_s$ are the reflection and transmition coefficients
respectively.
\section{The transformation law of the curvatures}
In the previous section, we summarize the perturbation equations
governing the physical modes.
On the extreme Reissner-Nordstr\"om background,
the quasinormal frequencies of the perturbations with the different
helicity coincide by the suitable shift of the total angular momentum
\cite{oursI,oursII}.
This fact suggests that the reflection and transmition amplitudes
are equivalent among the perturbations with the different helicity. 

It is well known that the extreme Reissner-Nordstr\"om background
has an unbroken global supersymmetry in $N=2$ supergravity
\cite{GibbonsHull}.
This implies that the perturbations with the different helicity
are related to each other.
 
In this section, we obtain the supersymmetric transformation laws 
between the curvatures of the perturbed fields through $N=2$ supergravity.
The action of the $N=2$ supergravity are represented by 
\begin{eqnarray}
	& &\Lag=-{\Mp^2 e \over 2}R-\half \eps^{\mu\nu\rho\sigma}
	(\psi^i_{A\mu}e^{A\dA}_\nu D_\rho \bpsi^i_{\dA\sigma}-
	-\bpsi^i_{\dA\mu} e^{A\dA}_\nu D_\rho \psi^i_{A\sigma} )
\nonumber \\
	& &\qquad -{i \over 32 \Mp^2} \eps^{\mu\nu\rho\sigma}~\Bigl[
	(\eps^{ij}\psi^{i A}_\mu \psi^j_{A\nu})~
	(\eps^{kl}\psi^{k B}_\rho \psi^l_{B\sigma}) 
	 -(\eps^{ij}\bpsi^{i\dA}_\mu \bpsi^j_{\dA\nu})~
	(\eps^{kl}\bpsi^{k\dB}_\rho \bpsi^l_{\dB\sigma})
	\Bigr]
\nonumber \\
	& &\qquad -{e \over 4}\hF_{\mu\nu}\hF^{\mu\nu}
	+{i \over 8 \Mp} \eps^{\mu\nu\rho\sigma} \hF_{\rho\sigma}
	\eps^{ij}~(\psi^{i A}_\mu \psi^j_{A\nu}
	+\bpsi^{i\dA}_\mu \bpsi^j_{\dA\nu} )~, 
\\
	& &\hF_{\mu\nu}= F_{\mu\nu}+{1 \over 2 \Mp}\eps^{ij}
	(\psi^{i A}_\mu \psi^j_{A\nu}-
	\bpsi^{i \dA}_\mu \bpsi^j_{\dA\nu}) \qquad (i,j,k,l=1,2)~,
\label{eq:lagrangian}
\end{eqnarray}
where $\Mp=(8\pi G)^{-1/2}$ is the Planck mass and 
$e^{A\dA}_\mu=e^a_\mu \sig^{A\dA}_a$, 
and $D_\mu$ is covariant derivative with respect to 
$\omega_\mu{}^{ab}$.
The connection $\omega_{\mu ab}$ including the torsion is given by 
tetrad and gravitini
through variating the action with respect to $\omega_{\mu ab}$,
\begin{mathletters}
\begin{eqnarray}
	& &\omega_\mu {}^{ab}=~\omega^{(0)}_\mu {}^{ab}+K_\mu {}^{ab}~,
\label{eq:defomega} \\
	& &\omega^{(0)}_\mu {}^{ab}=~e^{a\nu}\deri_{[\mu} e_{\nu]}^b
	-\half e_{c\mu}e^{a\nu}e^{b\lambda}\deri_\nu e^c_\lambda
	-(a \leftrightarrow b)~,
\label{eq:defomegazero} \\
	& &K_\mu {}^{ab}=~ {i \over 2 \Mp^2} \Bigl(e^{a\nu}
	\bpsi^i_{\dA [\mu}\sig^{b A\dA} \psi^i_{|A| \nu]}
	-\half e_{c\mu}e^{a\nu}e^{b\lambda} 
	\bpsi^i_{\dA \nu}\sig^{c A\dA} \psi^i_{A \lambda}
	\Bigr) 
\nonumber \\
	& & \qquad\qquad -(a \leftrightarrow b)~,
\label{eq:deftorsion}
\end{eqnarray}
\end{mathletters}
and the curvature is given by
\begin{mathletters}
\begin{eqnarray}
	R_{\mu\nu}{}^{ab}&=& ~2~ \deri_{[\mu}\omega_{\nu]}{}^{ab}
	+2~ \omega_{[\mu}{}^{ac} \omega_{\nu] c}{}^b~,
\label{eq:defcurvature} \\
	R_{ab}&= & ~e^\mu_a e^\nu_c R_{\mu\nu b}{}^c~.
\label{eq:defscalarcurvature}
\end{eqnarray}
\end{mathletters}
The action is invariant under the supersymmetric transformations,
\begin{mathletters}
\begin{eqnarray}
	& &\delta e_{a\mu}=~-{i \over 2\Mp}~
	(\alpha^i_A \sig^{A\dA}_a \bpsi^i_{\dA\mu}+
	\balpha^i_\dA \sig^{A\dA}_a \psi^i_{A\mu})~,
\label{eq:trantetrad} \\
	& &\delta A_\mu=~-\half \eps^{ij}~(\alpha^{i A}\psi^j_{A\mu}
	-\balpha^{i \dA}\bpsi^j_{\dA\mu})~,
\label{eq:tranA} \\
	&& \delta \psi^i_{A\mu}=~\Mp~ D_\mu \alpha^i_A
	-i~\eps^{ij}\hF_A{}^B~e_{\mu B\dA} \balpha^{j \dA}~,
\label{eq:tranRS}
\end{eqnarray}
\end{mathletters}
where $\alpha^i_A$ are Grassmann odd transformation parameters and
$\hF_{AB}$ is 2-spinor representation of the self-dual part of 
$\hF_{\mu\nu}$.

We can check that $\omega_{\mu ab}$ and 
$\hF_{\mu\nu}$ are supercovariant, $\ie$, their transformations
have no derivative of transformation parameters.
For the spin-$3/2$ fields, we introduce the supercovariant
curvatures of $\psi^i_{AB\dA}=e^\mu_{B\dA}\psi^i_{A\mu}$
in 2-spinor representation,
\begin{mathletters}
\begin{eqnarray}
	\Psi^{i A}{}_{B C}&=&\half \Bigl[ 
	D_{(B|\dA|} \psi^{i A}{}_{C)}{}^{\dA}
	+{i \over M_p} \eps^{ij} \hF^A {}_{(B} 
	\bpsi^{j \dA}{}_{|\dA| C)} \Bigr]~,
\label{eq:RScurspiI} \\
	\Psi^{i A}{}_{\dB \dC}&=&\half \Bigl[
	D_{B(\dB} \psi^{i AB}{}_{\dC)}
	-{i \over M_p} \eps^{ij} \hF^A {}_B
	\bpsi^j_{(\dB \dC)}{}^B \Bigr]~.
\label{eq:RScurspiII}
\end{eqnarray}
\end{mathletters}
They are transformed according to 
\begin{mathletters}
\begin{eqnarray}
	\delta \Psi^{i A}{}_{B C}&=&{\Mp \over 2}R_{BC}{}^{AD}~\alpha^i_D 
	+\halfi \eps^{ij}\Bigl[ D_{(B}{}^\dA~\hF_{C)}{}^A \Bigr]~\balpha^j_\dA
	+O(\psi^2)~,
\label{eq:RScurspitranI} \\
	\delta \Psi^{i A}{}_{\dB \dC}&=&
	{\Mp \over 2}R_{\dB\dC}{}^{AD}~\alpha^i_D 
	-\halfi \eps^{ij}\Bigl[ D_{B (\dB}~\hF^{A B} \Bigr]~\balpha^j_{\dC)}
\nonumber \\
	& &+{1 \over 2\Mp}~\hF^{AD}\bar \hF_{\dB\dC}~\alpha^i_D
	+O(\psi^2)~.
\label{eq:RScurspitranII}
\end{eqnarray}
\end{mathletters}
Because we will analyze the perturbations about 
a purely bosonic background,
it is sufficient to obtain the transformation laws 
at linear order of $\psi^i_{A\mu}$.

We introduce an expansion parameter $\lambda$ and 
replace fundamental fields, tetrad, connection, gravitini and 
electromagnetic potential about a background as, for example,
$\psi^i_{A\mu} \rightarrow \bfpsi^i_{A\mu}+\lambda \psi^i_{A\mu}$,
where we use the bold-face for the background quantities and
standard letters for the perturbed quantities, respectively.
Various equations and relations for perturbed fields are given by
expanding with respect to $\lambda$.

The perturbed fields have gauge degree of freedoms originated from
arbitrariness of correspondence between the perturbed world and
the background world.
Due to $\bfpsi^i_{A\mu}=0$, the bosonic quantities are invariant
under the supersymmetric gauge transformations and 
the fermionic quantities transform, for example, into
\begin{equation}
	\delta_g \Psi^i_{ABC}={\Mp \over 2}~\bfR_{BCA}{}^D \beta^i_D
	+\halfi \eps^{ij}\Bigl[\bfD_{(B}{}^\dA \bfF_{C)A} \Bigr]
	\bbeta^j_\dA~,
\label{eq:delgaugePsi}
\end{equation}
where $\beta^i_A$ are any spinor parameters.

Let us consider supersymmetric transformation laws.
Because of $\bfpsi^i_{A\mu}=0$,
the bosonic background quantities are invariant 
under supersymmetric transformation.
On the other hand, fermionic quantities generally change 
due to non-trivial bosonic background.
For example, the background gravitini transform 
under the supersymmetric transformation into
\begin{equation}
	\delta \bfpsi^i_{A\mu}=~\Mp~ \bfD_\mu \alpha^i_A
	-i \eps^{ij} \bfF_A {}^B~e_{\mu B\dA}~\balpha^{j \dA}~,
\label{eq:perturbtranRS} 
\end{equation}
and the perturbed supercovariant curvatures of gravitini 
transform into
\begin{eqnarray}
	& &\delta \Psi^i_{ABC}={\Mp \over 2}R_{BCA}{}^D~\alpha^i_D
	+\halfi \eps^{ij}
	\Bigl[\bfD_{(B}{}^\dA F_{C)A} \Bigr]~\balpha^j_\dA
\nonumber \\
	& &\qquad\qquad +\halfi \eps^{ij}\sig_{a(B}{}^\dA 
	\Bigl[\omega^a{}_{C)}{}^D \bfF_{DA}
	+\omega^a{}_{|A|}{}^D \bfF_{C)D}
	\Bigr]~\balpha^j_\dA~.
\label{eq:delperPsiI}
\end{eqnarray}

Therefore, if there are some supercovariantly constant spinors
(SCCS's),
\begin{equation}
	\bfD_\mu \zeta^i_A
	-{i \over \Mp}\eps^{ij} \bfF_A {}^B~
	e_{\mu B\dA}~\bzeta^{j \dA}=0~,
\label{eq:sccseq}
\end{equation}
the background configurations are invariant under the supersymmetric
transformations that are induced by SCCS's.
And then, unbroken supersymmetry persists on the system 
consisting of the perturbed fields.

We introduce the quantities constructed by $\Psi^i_{ABC}$, 
\begin{mathletters}
\begin{eqnarray}
	H^i_0 &\equiv & \Psi^i_{(ABC)}~o^A o^B o^C~,
\label{eq:defHo} \\
	H^i_1 &\equiv & \Psi^i_{(ABC)}~o^A o^B \iota^C~,
\label{eq:defHI} \\
	H^i_2 &\equiv & \Psi^i_{(ABC)}~o^A \iota^B \iota^C~,
\label{eq:defHII} \\
	H^i_3 &\equiv & \Psi^i_{(ABC)}~\iota^A \iota^B \iota^C~,
\label{eq:defHIII}
\end{eqnarray}
\end{mathletters}
where $o^A$ and $\iota^A$ are principal spinors of $\bfF_{AB}$.
Here we assume that the background spacetime is in the Petrov type D.
Then the physical modes are described by $H^i_0$ or $H^i_3$ 
because they are diffeomorphic, local Lorentz gauge invariant
due to the purely bosonic background and 
supersymmetric gauge invariant due to the type D character and 
Eq.(\ref{eq:delgaugePsi}).

Therefore we are interested in the transformation laws of
$H^i_0$ generated by SCCS's, $\zeta^i_A$,
\begin{eqnarray}
	\delta H^i_0 &=& {\Mp \over 2}
	\Bigl[\zeta^i_{(0)}\Psi_1- \zeta^i_{(1)}\Psi_0 \Bigr]
\nonumber \\
	&+& \halfi \eps^{ij}(\bfD_a \phi_0)
	(\bzeta^j_{(\dzero)}m^a-\bzeta^j_{(\done)}l^a)
\nonumber \\
	&+& i \eps^{ij}\phi_0(\bfbeta~ \bzeta^j_{(\dzero)}
	-\bfepsilon~ \bzeta^j_{(\done)})
\nonumber \\
	&-& i \eps^{ij}\bfphi_1~ 
	(\sig \bzeta^j_{(\dzero)}-\kappa \bzeta^j_{(\done)})~,
\label{eq:tranHo}
\end{eqnarray}
where $\zeta^i_{(0)}=o^A \zeta^i_A$ and 
$\zeta^i_{(1)}=\iota^A \zeta^i_A$ and they have the spin-weight
$+\half$ and $-\half$, respectively.
And then, $\Psi_0$, $\Psi_1$, $\phi_0$, $\sig$ and $\kappa$ are
perturbed Weyl scalars, Maxwell scalar and complex spin coefficients,
respectively.
Further $\bfepsilon$ and $\bfbeta$ are background spin coefficients.
\section{The relations of the reflection and 
	transmition coefficients}
In the previous section, we obtained the transformation law between
the perturbed curvatures of gravitini.
Using it, we can relate the decoupled modes $Y_{+s}$ on
the extreme Reissner-Nordstr\"om black hole.

On the extreme hole, there exist the supercovariantly constant
spinors,
\begin{mathletters}
\begin{eqnarray}
	& &\zeta^i_{(0)}=\sqrt{2}~\eta^i_{(0)}(\theta)\exp(i \dm \phi)~,
\label{eq:zetazero} \\
	& &\zeta^i_{(1)}={\Delta^\half \over r}
	~\eta^i_{(1)}(\theta) \exp(i \dm \phi)~,
\label{eq:zetaI}
\end{eqnarray}
\end{mathletters}
where $\dm$ is $+\half$ or $-\half$ and $\eta^i_A$ satisfy
\begin{mathletters}
\begin{eqnarray}
	& &\Lag^{\dm\dagger}_{-\half} \eta^i_{(0)}
	=\Lag^{\dm}_{-\half} \eta^i_{(1)}=0~,
\\
	& &\Lag^{\dm}_{+\half} \eta^i_{(0)}=\eta^i_{(1)}~,
\\
	& &\Lag^{\dm\dagger}_{+\half} \eta^i_{(1)}=-\eta^i_{(0)}~,
\label{eq:etaformula}
\end{eqnarray}
\end{mathletters}
where the operators $\Lag^m_n$ and $\Lag^{m\dagger}_n$ are the same
operators as defined in Eqs.(\ref{mlet:defLag})
in the previous section and we manifest 
azimuthal angular momentum dependence with index $m$.
The supercovariantly constant spinors satisfy the relation,
\begin{equation}
	{i \over \Mp}\eps^{ij}\bfphi_1 \bzeta^j_\dA=
	-\bfgamma \Bigl( {2 r^2 \over \Delta} \biota_\dA \iota^A
	+\bo_\dA o^A \Bigr) \zeta^i_A~.
\label{eq:zetarel}
\end{equation}

From Eq.(\ref{eq:zetarel}), the transformation of $H^i_0$, 
Eq.(\ref{eq:tranHo}) becomes
\begin{equation}
	\delta H^i_0={\Mp \over 2}
	\Bigl[{\Delta^\half \over r} 
	\Bigl( \Psi_0+4 \bfgamma{r^2 \over \Delta}\sigma \Bigr)
	\eta^i_{(1)}- \sqrt{2}(\Psi_1-2\bfgamma\kappa)~
	\eta^i_{(0)}\Bigr]~e^{i \dm \phi}~,
\label{eq:tranHophantomI}
\end{equation}
where we, of course adopt the phantom gauge $\phi_0=\phi_2=0$.

According to Sec. II, we decompose $\Psi_0$, $\Psi_1$,
$\kappa$ and $\sigma$ by the spin-weighted spherical harmonics,
and we manifest angular momentum dependence.
For example,
\begin{eqnarray}
	& &\Psi_0=R^J_{+2}(r)S^J_{+2}(\theta) e^{i(\Omega t+m \phi)}~,
\\
	& &\Lag^{m\dagger}_{-1}\Lag^m_2  S^J_{+2}=-\mu_J^2 S^J_{+2}~,
\\
	& &\mu_J=\sqrt{(J-1)(J+2)}~.
\label{eq:sphericalharmonics}
\end{eqnarray}
And then 
\begin{eqnarray}
	\delta H^i_0&=&{\Mp \over 2}
	\Bigl[{\Delta^\half \over r} 
	\Bigl( R^J_{+2}+r {d \over dr}\ln 
	\Bigl({\Delta \over r^2} \Bigr) s^J \Bigr)
	(\eta^i_{(1)}S^J_{+2})
\nonumber \\
	& &\qquad -{1 \over r} \Bigl( R^J_{+1}
	-r^3 {d \over dr}\Bigl({\Delta \over r^2} \Bigr) k^J
	\Bigr)~	(\eta^i_{(0)} S^J_{+1}) \Bigr] 
	e^{i (\Omega t+M \phi)}~,
\label{eq:tranHophantomII}
\end{eqnarray}
where $M=m+\dm$.

The each quantity, $\eta^i_{(1)} S^J_{+2}$ and $\eta^i_{(0)} S^J_{+1}$,
has spin-weight $+3/2$, 
but not an eigenstate of total angular momentum respectively.
Hence we need decompose them into ($S^{J+1/2}_{+3/2}$) and 
($S^{J-1/2}_{+3/2}$) that are eigenstates of the total angular momentum.
It is easy to check the equations
\begin{mathletters}
\begin{eqnarray}
	& &\Lag^{M\dagger}_{-\half}\Lag^M_{\thirdh}
	(\eta^i_{(1)}S^J_{+2})
	=-\mu_J (\eta^i_{(0)}S^J_{+1})
	-\mu_J^2 (\eta^i_{(1)}S^J_{+2})~,
\label{eq:eqI} \\
	& &\Lag^{M\dagger}_{-\half}\Lag^M_{\thirdh}
	(\eta^i_{(0)}S^J_{+1})
	=-\mu_J (\eta^i_{(1)}S^J_{+2})
	-(\mu_J^2+3) (\eta^i_{(0)}S^J_{+1})~.
\label{eq:eqII}
\end{eqnarray}
\end{mathletters}
From these equations, we can decompose $(\eta^i_{(1)} S^J_{+2})$ and 
$(\eta^i_{(0)} S^J_{+1})$ as 
\begin{mathletters}
\begin{eqnarray}
	\eta^i_{(1)}S^J_{+2}&=&\xi^i {S^{J+\half}_{+\thirdh}
	-S^{J-\half}_{+\thirdh} \over q_2-q_1}~,
\\
	\eta^i_{(0)}S^J_{+1}&=&\xi^i {{q_1}~ S^{J+\half}_{+\thirdh}
	- {q_2}~ S^{J-\half}_{+\thirdh} \over 2 \mu_J (q_2-q_1) }~,
\label{eq:decompose}
\end{eqnarray}
\end{mathletters}
where $\xi^i$ are arbitrary Grassmann odd constants and 
$q_{1,2}$ are defined by
\begin{eqnarray}
	q_1 &\equiv & 3 + \sqrt{9+4 \mu_J^2}=2 (J+2)~,
\nonumber \\
	q_2 &\equiv & 3 - \sqrt{9+4 \mu_J^2}= - 2 (J-1)~.
\label{eq:defqII}
\end{eqnarray}
The expressions of $q_j$ are the extreme limit of Eq.(\ref{eq:defqI}).

Because $\Delta=(r-1)^2$ in the extreme case, 
Eq.(\ref{eq:tranHophantomII}) is rewritten as
\begin{eqnarray}
	\delta H^i_0 &= &{\Mp \Delta^{\half} \xi^i \over 
	2 r(q_2-q_1)} 
	\Bigl[ S^{J+\half}_{+\thirdh} \Bigl(
	F^J_{+1}-{q_1 \over 2 \Delta^\half \mu_J}G^J_{+2}
	\Bigr)
\nonumber \\
	& & \quad\qquad\qquad  - S^{J-\half}_{+\thirdh} \Bigl(
	F^J_{+2}-{q_2 \over 2 \Delta^\half \mu_J}G^J_{+1}
	\Bigr) \Bigr] e^{i(\Omega t+M \phi)}~,
\nonumber \\
	&=&{\Mp \Delta^{\half} \xi^i \over 2 r(q_2-q_1)}\Bigl[
	S^{J+\half}_{+\thirdh} \Bigl\{
	F^J_{+1}-{q_1 \over 2}
	{r \Delta^\half \over \mu_J^2 r+2 q_1}\Bigl(
	D^\dagger_2-{3 \over r} \Bigr)F^J_{+1}
	\Bigr\}
\nonumber \\
	& & \qquad - S^{J-\half}_{+\thirdh} \Bigl\{
	F^J_{+2}-{q_2 \over 2}
	{r \Delta^\half \over \mu_J^2 r+2 q_2}\Bigl(
	D^\dagger_2-{3 \over r} \Bigr)F^J_{+2}
	\Bigr\} \Bigr] e^{i(\Omega t+M \phi)}~,
\label{eq:tranHophantomIII}
\end{eqnarray}
where we use the relation Eq.(\ref{eq:relGF}).
Eq.(\ref{eq:tranHophantomIII}) shows that 
the helicity-($+3/2$) modes with $J+1/2$ are generated by 
the unbroken supersymmetry from the helicity-($+1$) mode 
$F^J_{+1}$ with total angular momentum $J$
or the helicity-($+2$) mode $F^{J+1}_{+2}$ with $J+1$.

From Eq.(\ref{eq:tranHophantomIII}), radial parts $Y^{J_s}_{+\thirdh}$
of perturbed curvatures of gravitini are generated from $Y^J_{+s}$ as
\begin{mathletters}
\label{mlet:relationY}
\begin{equation}
	Y^{k J_s}_{+\thirdh}= \xi^k \Bigl[ Y^J_{+s}
	-C^J_s(r) ~\Lambda_- Y^J_{+s} \Bigr]~,
\label{eq:relY}
\end{equation}
or equivalently,
\begin{equation}
	\xi^k \Bigl[ 2 i \Omega +{1 \over C^J_s }-C^J_s Q_s
	+\half P_{\thirdh} \Bigr] Y_{+s}^J 
	=\Bigl[ 2 i \Omega +{1 \over C^J_s }
	+\half P_{\thirdh} \Bigr] Y_{+\thirdh}^{k J_s}
	+\Lambda_- Y_{+\thirdh}^{k J_s} ~,
\label{eq:relinvY} \\
\end{equation}
\end{mathletters}
where
\begin{equation}
	C^J_s (r) \equiv 
	{q_s r^2 \Delta^{\thirdh} \over 2 \mu^2_J D_s}~,
\qquad\qquad
	D_s \equiv \Delta^2 
	\Bigl( 1+ {2 q_s \over \mu^2_J r} \Bigr)~,
\label{eq:defCsjDs}
\end{equation}
and $J_s$ is $J+\half$ for $s=1$ and $J-\half$ for $s=2$.
Eqs.(\ref{mlet:relationY})
are our main result and
in principle, we can also obtain the relations between potentials
of perturbations with different helicities.
Hereafter we omit the index $k$ which distinguishes two gravitini.

From Eqs.(\ref{mlet:relationY}), 
we can obtain the relation between
reflection and transmition amplitudes.
From them, it follows that 
$Y^{J_s}_{+\thirdh}$ derived 
from $Y^{J \pinfin}_{+s}$ and $Y^{J \pinfout}_{+s}$($s=1,2$)
have, respectively, the asymptotic behaviors 
at $r_* \rightarrow \infty$
\begin{equation}
	Y^{J_s}_{+\thirdh} \sim Y^{J_s \pinfin}_{+\thirdh}
\qquad \hbox{and} \qquad
	Y^{J_s}_{+\thirdh} \sim 
	{i \Omega q_s \over \mu^2_J} {K_s \over K_{\thirdh}}
	Y^{J_s \pinfout}_{+\thirdh}~.
\label{eq:asymgYthirdhp}
\end{equation}
Similarly, it follows that $Y^{J_s}_{+\thirdh}$ derived 
from $Y^{J \minfin}_{+s}$ and $Y^{J \minfout}_{+s}$($s=1,2$)
have, respectively, the asymptotic behaviors
at $r_* \rightarrow -\infty$ 
\begin{equation}
	Y^{J_s}_{+\thirdh} \sim Y^{J_s \minfout}_{+\thirdh}
\qquad \hbox{and} \qquad
	Y^{J_s}_{+\thirdh} \sim 
	{i \Omega q_s \over \mu^2_J} {K_s \over K_{\thirdh}}
	Y^{J_s \minfin}_{+\thirdh}~.
\label{eq:asymgYthirdhm}
\end{equation}

Therefore the asymptotic form of $Y^{J_s}_{+\thirdh}$ derived 
from the solution for $Y^J_{+s}$($s=1,2$) 
having the asymptotic behavior
\begin{eqnarray}
	Y^J_{+s} &\sim& Y^{J \pinfin}_{+s} 
	+ R^J_s (\Omega) Y^{J \pinfout}_{+s} 
		\qquad (r_* \rightarrow \infty)~,
\nonumber \\
	& \sim &\qquad\qquad\qquad T^J_s (\Omega) Y^{J \minfout}_{+s} 
		\qquad (r_* \rightarrow -\infty)~,
\label{eq:asymYJs}
\end{eqnarray}
has the asymptotic behavior
\begin{eqnarray}
	Y^{J_s}_{+\thirdh} & \sim & Y^{J_s \pinfin}_{+\thirdh}
	+ R^J_s (\Omega) {i \Omega q_s \over \mu^2_J}
	{K_s \over K_{\thirdh}} Y^{J_s \pinfout}_{+\thirdh}
		\qquad (r_* \rightarrow \infty)~,
\nonumber \\
	& \sim & \qquad\qquad\qquad 
	T^J_s (\Omega) Y^{J_s \minfout}_{+\thirdh}
		\qquad\qquad\qquad (r_* \rightarrow -\infty)~.
\label{eq:asymYthird}
\end{eqnarray}
Accordingly, we obtain the relations of reflection and transmition 
coefficients, 
\begin{eqnarray}
	& &R^{J_s}_{\thirdh} (\Omega) = \gamma_s R^J_s (\Omega) 
\qquad
	T^{J_s}_{\thirdh} (\Omega) = T^J_s (\Omega) \qquad (s=1,2)~,
\\
	& &\gamma_s \equiv {i \Omega q_s \over \mu^2_J}
	{K_s \over K_{\thirdh}}~,
\label{eq:relationcoeff}
\end{eqnarray}
where $|\gamma_s|=1$.
Thus, under the suitable shift of angular momentums,
while the amplitudes of the transmitted waves are identically
the same for three perturbed fields, the reflected amplitudes differ
only in their phases.
\section{Summary}
In the previous section, using the unbroken supersymmetry that
remains on the extreme Reissner-Nordstr\"om black hole, 
we obtain the relation between the reflection and 
transmition coefficients of decoupled modes with 
(helicity, total angular momentum)
=($1,J$), ($\thirdh,J+\half$), ($2,J+1$).

These relations are also expected for the perturbations about
the superpartners of the extreme Reissner-Nordstr\"om
black hole 
\cite{AichelburgEmbacher}
and for matter multiplets about them.

In the previous paper
\cite{oursII},
we observed that the Regge-Wheeler potential of 
gravitational perturbation coincides with one of 
electromagnetic perturbation by inversion of the tortoise coordinate,
that is, exchange of the horizon for infinity, vice versa.
It is interesting to understand the above correspondence by
using the relations of the perturbations obtained 
in the previous section.
\acknowledgments
The author would like to thanks H. Onozawa, T. Mishima 
and H. Ishihara for valuable comments and stimulating discussions.
He also appreciate Professor A. Hosoya for 
continuous encouragement.
The research was supported in part by 
the Scientific Research Fund of the Ministry of Education.


\begin{thebibliography}{99}
\bibitem{WittenOlive}
	E. Witten and D. Olive, \pl{\bf 78B}, 97 (1978)~.
\bibitem{Hajicek}
	P. Hajicek, Nucl.Phys. {\bf B185}, 254 (1981)~.
\bibitem{Gibbons}
	G.W. Gibbons, in {\it Supersymmetry, Supergravity 
	and Related Topics}, proceedings of the XV th GIFT Seminar,
	edited by F. Augila, et.al. (World Scientific, 1985)~.
\bibitem{algebra}
	C. Teitelboim, \pl{\bf 69B}, 240 (1977)~.
\bibitem{positiveEsugra}
	S. Deser and C. Teitelboim, \prl{\bf 39}, 249 (1977)~;
	M. Grisaru, \pl{\bf 37B}, 249 (1978)~.
\bibitem{positiveEeinstein}
	E. Witten, Commun.Math.Phys.{\bf 80}, 381 (1981)~;
	J. M. Nester, \pl{\bf 83A}, 241 (1981)~.
\bibitem{GibbonsHull}
	G. W. Gibbons and C. M. Hull, \pl{\bf 109B}, 190 (1982)~.
\bibitem{KalloshLindeETAL}
	R. Kallosh, A. Linde, T. Ortin and A. Peet, 
	\prd{\bf 46}, 5278 (1992)~.
\bibitem{GibbonsKastorETAL}
	G. W. Gibbons, D. Kastor, L. A. J. London, P. K. Townsend
	and J. Traschen,
	Nucl.Phys.{\bf B416}, 850 (1994)~.
\bibitem{Kallosh}
	R. Kallosh, \pl{\bf 282B}, 80 (1992)~.
\bibitem{AichelburgGuven}
	P. Cordero and C. Teitelboim, \pl{\bf 78B}, 80 (1978)~;
	R. G\"uven, {\it ibid.}\ {\bf 22}, 2327 (1980)~;
	P.C. Aichelburg and R. G\"uven, {\it ibid.}\ {\bf 24}, 2066 (1981)~;
	R. G\"uven, {\it ibid.}\ {\bf 25}, 3117 (1982)~;
	P.C. Aichelburg and R. G\"uven, {\it ibid.}\ {\bf 27}, 456 (1983)~.
\bibitem{HolzheyWilczek}
	C.F.E. Holzhey and F. Wilczek, Nucl.Phys.{\bf B380}, 447 (1992)~.
\bibitem{oursI}
	H. Onozawa, T. Mishima, T. Okamura and H. Ishihara,
	\prd{\bf 53}, 7033 (1996)~.
\bibitem{oursII}
	H. Onozawa, T. Okamura, T. Mishima and H. Ishihara,
	\prd{\bf 55}, (1997)~.
\bibitem{sugra}
	S. Ferrara and P. van Nieuwenhuizen, \prl{\bf 37}, 1669 (1976)~.
\bibitem{Chandra}
	S. Chandrasekhar, {\it The Mathematical Theory of 
	Black Holes}, (Clarendon, Oxford, 1983)~.
\bibitem{TorresETAL}
	G. F. Torres del Castillo and G. Silva-Ortigoza,
	\prd{\bf 46}, 5395 (1992)~.
\bibitem{phantom}
	See the \cite{Chandra}~, pp240~.
\bibitem{AichelburgEmbacher}
	P. C. Aichelburg and F. Embacher, \prd{\bf 34}, 3006 (1986)~.

\end{thebibliography}
\end{document}